\documentclass[a4paper]{jpconf}
\usepackage{graphicx}
\begin{document}
\title{Pseudospin S=1 description of the cuprate complexity: the charge triplet’s model}

\author{A.S. Moskvin}

\address{Department of Theoretical Physics, Ural Federal University, 620083 Ekaterinburg,  Russia}

\ead{alexander.moskvin@urfu.ru}

\begin{abstract}
We introduce a minimal model for 2D cuprates  with the on-site Hilbert space  reduced to only three effective valence centers CuO$_4^{7-,6-,5-}$ (nominally Cu$^{1+,2+,3+}$) and make use of the S=1 pseudospin formalism. Despite its seeming simplicity the model is believed to capture the salient features both of the hole- and electron-doped cuprates. The pseudospin formalism elucidates an unique fermion-boson duality of the doped cuprates, does provide an unified standpoint for classification of the "myriad"\, of  electronic phases in cuprates and the evolution of the CuO$_2$ planes under a nonisovalent doping, introduces the on-site mixed valence quantum superpositions and order parameters to be novel features of the cuprate physics, does provide a comprehensive description of the correlated one- and two-particle transport, coexistence of $p$- and $n$-type carriers, electron-hole asymmetry,  anticorrelation of conventional spin and superconducting order parameters.  Concept of the electron and hole centers, differing by a composite local boson, and electron-hole pairing are shown to explain central points of the cuprate puzzles, in particular, the HTSC itself, pseudogap, and Fermi surface reconstruction.
\end{abstract}

\section{Introduction}
Recently\,[1] we argued that an unique property of high-T$_c$ cuprates is related with a dual nature of the Mott insulating state of the parent compounds that manifests itself in two distinct energy scales for the charge transfer (CT) reaction: Cu$^{2+}$\,+\,Cu$^{2+}$\,$\rightarrow$\,Cu$^{1+}$\,+\,Cu$^{3+}$.
Indeed, the $d$\,-\,$d$ CT gap as derived from the optical measurements in parent cuprates such as La$_2$CuO$_4$ is 1.5-2.0\,eV while  the true (thermal) $d$\,-\,$d$ CT gap, or effective correlation  parameter $U_d$,  appears to be as small as 0.4-0.5\,eV. 
It means cuprates should be addressed to be \emph{d-d} CT unstable systems whose description implies accounting of the three many-electron valence states CuO$_4^{7-,6-,5-}$ (Cu$^{1+,2+,3+}$) on an equal footing as a well-defined charge triplet.

Below we introduce a minimal model  with the on site Hilbert space  reduced to only three states, three effective valence centers CuO$_4^{7-,6-,5-}$ (nominally Cu$^{1+,2+,3+}$) where the electronic and lattice degrees of freedom get strongly locked together, and make use of
the S=1 pseudospin formalism we have proposed earlier\,[1-4]. Such a formalism constitutes a powerful method to study complex phenomena in interacting quantum systems characterized
by the coexistence and competition of various ordered states\,[5].

\section{Pseudospin formalism}
In frames of the pseudospin formalism we address  a mixed-valence (MV) system with three possible stable on-site valence states $M^0, M^{\pm}$, hereafter a bare center $M^0$=CuO$_4^{6-}$, a hole center $M^{+}$=CuO$_4^{5-}$, and an electron center $M^{-}$=CuO$_4^{7-}$, respectively, 
 and neglect all other degrees of freedom focusing on the quantum charges.  Three different valence charge states $M^0,M^{\pm}$ we associate with three components of the $S=1$ pseudo-spin (isospin) triplet with  $M_S =0,\pm 1$, respectively.
Such a correspondence immediately implies introduction of the unconventional on-site  MV  quantum superpositions $|\Psi\rangle =\sum_{M}c_M|1M\rangle $, or
\begin{equation}
	|\Psi\rangle = c_{-1}|Cu^{1+}\rangle + c_{0}|Cu^{2+}\rangle + c_{1}|Cu^{3+}\rangle  \, ,
	\label{site} 
\end{equation}
that points to many novel effects related with local Cu states. 
However, we should note that at variance with spinless ground states of Cu$^{1+}$  and Cu$^{3+}$ centers the bare Cu$^{2+}$ center has a conventional spin s=1/2, in other words we arrive at the S\,=\,1 pseudospin model with doubly degenerate M=0 value.
In the partition function of the classical spin model, this leads to a factor of 2 for every Cu$^{2+}$ site.

The $S=1$ spin algebra includes eight independent nontrivial pseudo-spin operators, three dipole and five quadrupole operators:
\begin{equation}
	{\hat S}_z; {\hat S}_{\pm}=\mp \frac{1}{\sqrt{2}}(S_{x}\pm iS_{y});{\hat S}_z^2;{\hat T}_{\pm}=\{S_z, S_{\pm}\};{\hat S}^2_{\pm}\, .
\end{equation}
The two fermion-like  pseudospin raising/lowering operators
$S_{\pm}$ and $T_{\pm}$  change the pseudo-spin projection by $\pm 1$, with slightly different properties
$$
\langle 0 |\hat S_{\pm} | \mp 1 \rangle = \langle \pm 1 |\hat S_{\pm} | 0
\rangle =\mp 1 \,,
$$
\begin{equation}
\langle 0 |\hat T_{\pm}| \mp 1 \rangle = -\langle \pm 1 |(\hat T_{\pm}| 0 \rangle =+1. \label{S2}
\end{equation}
In lieu of ${\hat S}_{\pm}$ and ${\hat T}_{\pm}$ operators one may use  two novel operators ${\hat P}_{\pm}$ and ${\hat N}_{\pm}$:
\begin{equation}
	{\hat P}_{\pm}=\frac{1}{2}({\hat S}_{\pm}+{\hat T}_{\pm});\,{\hat N}_{\pm}=\frac{1}{2}({\hat S}_{\pm}-{\hat T}_{\pm})\,, 
\end{equation} 
which do realize transformations Cu$^{2+}$$\leftrightarrow$Cu$^{3+}$ and Cu$^{1+}$$\leftrightarrow$Cu$^{2+}$, respectively.
The  boson-like  pseudospin raising/lowering operators
${\hat S}_{\pm}^{2}$ do
change the pseudo-spin projection by $\pm 2$ and define a local nematic order parameter $\langle S_{\pm}^{2} \rangle\,=\,\frac{1}{2}(\langle S_x^2-S_y^2\rangle \pm i\langle\{S_x,S_y\}\rangle )$.
This on-site off-diagonal order parameter is nonzero only for the on-site $M^-$(Cu$^{1+}$)-$M^+$(Cu$^{3+}$) superpositions. It is worth noting that the ${\hat S}_{+}^{2}$ (${\hat S}_{-}^{2}$) operator
creates an on-site hole (electron) pair, or composite boson, with a kinematic constraint $({\hat S}_{\pm}^{2})^2$\,=\,0, that underlines its "hard-core"\, nature. Obviously, the pseudospin nematic average $\langle S_{\pm}^{2} \rangle$ can be addressed to be a local complex superconducting order parameter.
Both ${\hat S}_{+}$(${\hat S}_{-}$) and ${\hat T}_{+}$(${\hat T}_{-}$) can be anyhow related with a conventional s=1/2 spin single particle creation (annihilation) operators, however, these are not standard fermionic ones, as well as ${\hat S}_{+}^2$(${\hat S}_{-}^2$) operators are not standard bosonic ones.

\section{Effective S=1 pseudospin Hamiltonian}
Effective S=1 pseudospin Hamiltonian which does commute with the z-component of the total pseudospin  $\sum_{i}S_{iz}$ thus conserving the total charge of the system  can be written to be a sum of potential and kinetic energies: 
\begin{equation}
	{\hat H}={\hat H}_{ch}+{\hat H}_{tr}\,,
\label{H}	
\end{equation}
where
\begin{equation}
	{\hat H}_{ch} =  \sum_{i}  (\Delta _{i}S_{iz}^2
	  - (\mu -h_{i})S_{iz}) + \sum_{i<j} V_{ij}S_{iz}S_{jz}\, ,
\label{Hch}	   
\end{equation}
and ${\hat H}_{tr}={\hat H}_{tr}^{(1)}+{\hat H}_{tr}^{(2)}$ being a sum of one-particle and two-particle transfer contributions. In terms of ${\hat S}_{\pm}$ and  ${\hat T}_{\pm}$ operators ${\hat H}_{tr}^{(1)}$ and ${\hat H}_{tr}^{(2)}$ read as follows:
$$ 
{\hat H}_{tr}^{(1)}= \sum_{i<j} t_{ij}(S_{i+}S_{j-}+S_{i-}S_{j+})+
 \sum_{i<j} t_{ij}^{\prime}(T_{i+}T_{j-}+T_{i-}T_{j+})
$$ 
\begin{equation} 
 +\frac{1}{2}\sum_{i<j} t_{ij}^{\prime\prime}(S_{i+}T_{j-}+S_{i-}T_{j+}+T_{i+}S_{j-}+T_{i-}S_{j+})\, ;
\label{H-ST}	
\end{equation}
\begin{equation}
  {\hat H}_{tr}^{(2)}=\sum_{i<j} t_{ij}^b(S_{i+}^{2}S_{j-}^{2}+S_{i-}^{2}S_{j+}^{2})\,,
  \label{H2}
\end{equation}
with a charge density constraint: 
\begin{equation}
\frac{1}{2N}\sum _{i} \langle S_{iz}\rangle =\Delta n	\, ,
\label{Sz}
\end{equation}
where $\Delta n$ is the deviation from a half-filling ($N_{M^+}$\,=\,$N_{M^-}$). 
Hamiltonian ${\hat H}_{ch}$ corresponds to a classical spin-1 Ising model with a single-ion anisotropy in the presence of a longitudinal magnetic field. 
The first single-site term describes the effects of a bare pseudo-spin splitting, or the local energy of $M^{0,\pm}$ centers and relates with the on-site density-density interactions. The second term   may be
related to a   pseudo-magnetic field ${\bf h}_i$\,$\parallel$\,$Z$, in particular, a real electric field which acts as a chemical potential ($\mu$ is the hole chemical potential, and $h_i$ is a (random) site energy).  The third term in ${\hat H}_{ch}$ describes the effects of the short-  and long-range inter-site density-density interactions including  screened Coulomb and covalent couplings.

Hamiltonian ${\hat H}_{tr}^{(1)}$  describes the one-particle inter-site  hopping and  represents an obvious extension of the conventional Hubbard model which assumes that the electronic orbital is infinitely rigid irrespective of occupation number, and has much in common with so-called dynamic Hubbard models\,[6] that describe a correlated hopping.  Indeed, conventional Hubbard model implies all the single particle transfer (\ref{H-ST}) is governed only by the $SS$-term in (\ref{H-ST}) while the $TT$ and $ST$ terms describe a non-Hubbard correlated density-dependent  single-particle hopping.
In terms of ${\hat P}_{\pm}$ and  ${\hat N}_{\pm}$ operators the  Hamiltonian ${\hat H}_{tr}^{(1)}$ transforms as follows: 
$$ 
{\hat H}_{tr}^{(1)}= \sum_{i<j} t^p_{ij}(P_{i+}P_{j-}+P_{i-}P_{j+})+
 \sum_{i<j} t^n_{ij}(N_{i+}N_{j-}+N_{i-}N_{j+})
$$ 
\begin{equation} 
 +\frac{1}{2}\sum_{i<j} t^{pn}_{ij}(P_{i+}N_{j-}+P_{i-}N_{j+}+N_{i+}P_{j-}+N_{i-}P_{j+})\, ,
\label{H1a}	
\end{equation}
where
\begin{equation}
	t^{p,n}_{ij}=t_{ij}+t_{ij}^{\prime}\pm t_{ij}^{\prime\prime};\, 
	t^{pn}_{ij}=t_{ij}-t_{ij}^{\prime}\, .
\end{equation}
All the three terms in (\ref{H1a})  suppose a clear physical interpretation. The first $PP$-type term describes one-particle hopping processes: 
$Cu^{3+}+Cu^{2+}\leftrightarrow Cu^{2+}+Cu^{3+}$,
that is  a rather conventional  motion of the hole $M^+$ ($Cu^{3+}$) centers in the lattice formed by $M^0$ ($Cu^{2+}$)-centers ($p$-type carriers, respectively) or the motion of the $M^0$ ($Cu^{2+}$)-centers in the lattice formed by hole $M^+$ ($Cu^{3+}$) centers ($n$-type  carriers, respectively). 
The second $NN$-type term describes 
one-particle hopping processes: $Cu^{1+}+Cu^{2+}\leftrightarrow Cu^{2+}+Cu^{1+}$,
that is  a rather conventional  motion of the electron $M^-$ ($Cu^{1+}$) centers in the lattice formed by $M^0$ ($Cu^{2+}$)-centers ($n$-type carriers) or the motion of the $M^0$ ($Cu^{2+}$)-centers in the lattice formed by electron $M^-$ ($Cu^{1+}$) centers ($p$-type  carriers).
These hopping processes are typical ones for heavily underdoped or heavily overdoped cuprates. 
It is worth noting that the ST-type contribution of the one-particle transfer differs in sign for the $PP$ and $NN$ transfer thus breaking the electron-hole symmetry.
The third $PN$ ($NP$) term in (\ref{H1a}) defines a very different one-particle hopping process:
$Cu^{2+}+Cu^{2+}\leftrightarrow Cu^{3+}+Cu^{1+}, Cu^{1+}+Cu^{3+}$,
that is the {\it local disproportionation/recombination}, or the {\it $EH$-pair creation/annihilation}. Interestingly, the term can be related with a local pairing as the $Cu^{1+}$ center can be addressed to be electron pair (= composite electron boson) localized on the  $Cu^{3+}$ center or {\it vice versa} the $Cu^{3+}$ center can be addressed to be hole pair (= composite hole boson) localized on the  $Cu^{1+}$ center. 
 
Hamiltonian ${\hat H}_{tr}^{(2)}$ describes the two-particle (local composite boson) inter-site  hopping, that is the  motion of the electron (hole) center in the lattice formed by the hole (electron) centers, or the exchange reaction:
$Cu^{3+}+Cu^{1+} \leftrightarrow Cu^{1+}+Cu^{3+}$.
  In other words, $t^b_{ij}$ is the transfer integral for the local composite boson.

Conventional spin degree of freedom can be build in our effective Hamiltonian, if we take into account Heisenberg spin exchange Cu$^{2+}$-Cu$^{2+}$ coupling as follows
\begin{equation}
	{\hat H}_{ex}=\sum_{i>j}(1-{\hat S}_{iz}^2)(1-{\hat S}_{jz}^2)I_{ij}(\hat {\bf s}_i\cdot \hat {\bf s}_j)\, ,
	\label{Hspin}
\end{equation}
where $(1-{\hat S}_{iz}^2)$ is a projection operator which picks out the s=1/2 Cu$^{2+}$ center, $I_{ij}$ is an exchange integral. 
Obviously, the spin exchange provides an energy gain to the parent antiferromagnetic insulating (AFMI) phase with $\langle {\hat S}_{iz}^2\rangle$\,=\,0, while local superconducting order parameter is maximal given $\langle {\hat S}_{iz}^2\rangle$\,=\,1. In other words, the superconductivity and magnetism are nonsymbiotic phenomena with competing order parameters giving rise to an inter-twinning, glassiness, and other forms of electronic heterogeneities. 

Effective pseudospin Hamiltonian (\ref{H}) is significantly complicated as compared with  
a typical S\,=\,1 spin Hamiltonian with uniaxial single-ion and exchange anisotropies on a square lattice:
\begin{equation}
	{\hat H}=\sum_{i>j}J_{ij}(S_{ix}S_{jx}+S_{iy}S_{jy}+\lambda S_{iz}S_{jz})+\sum_iDS_{iz}^2 - \sum_ihS_{iz}\, ,
	\label{Hs}
\end{equation}
investigated rather extensively in recent years (see, e.g., Refs.\,[7] and references therein).      
Correspondence with our pseudospin Hamiltonian points to $D=\Delta$, $J_{ij}=-t_{ij}$, $\lambda J_{ij}=V_{ij}$ ($t_{ij}^{\prime}=t_{ij}^{\prime\prime}$\,=\,0; $t_{ij}^{b}$\,=\,0).
The Hamiltonian (\ref{Hs}) is invariant under the transformation $J,\lambda\rightarrow -J,-\lambda$ and a shift of the Brillouin zone ${\bf k}\rightarrow {\bf k}+(\pi ,\pi )$. 
The spectrum of the spin Hamiltonian (\ref{Hs}) in the absence of external magnetic field changes drastically as $\Delta$
varies from very small to very large positive or negative values. A strong "easy-plane" anisotropy for large positive $\Delta >0$ favors a singlet phase where all the spins are in the $S_z = 0$   ground state. This quadrupole ($Q_{zz}$\,=\,-$\frac{2}{3}\rangle$) phase has no magnetic order, and is aptly referred to as a quantum paramagnetic phase (QPM), which is separated
from the "ordered" state by a quantum critical point (QCP) at some $\Delta$\,=\,$\Delta_{1}$. 
A strong "easy-axis" anisotropy for large negative $\Delta \leq\Delta_2$, favors a spin ordering along $Z$, the "easy axis", with the on-site $S_z = \pm 1$ ($Z$-phase). The order parameter will be "Ising-like"\, and long-range (staggered) diagonal order will persist at finite temperature, up to a critical line T$_c(\Delta )$.   
For intermediate values $\Delta_1>\Delta >\Delta_2$ the Hamiltonian will have O(2) symmetry and the system is in a
gapless $XY$ phase. At T\,=\,0 the  O(2) symmetry will be spontaneously broken and the system will exhibit spin order in some direction.  Although there will be no ordered phase at finite temperature one expects a finite temperature Kosterlitz-Thouless transition. 
 At finite effective field $h_z$ but $\lambda$\,=\,1 the $XY$ phase transforms into a canted antiferromagnetic $XY$-$Z_{FM}$ phase, the spins acquire a uniform longitudinal component which increases with field and saturates at the fully polarized
(FP) state (all $S_z$\,=\,1, $Z_{FM}$ phase) above the saturation field $h_s$. 
However, at $D>$\,0  and $\lambda >$\,1 the phase diagram contains an extended spin supersolid or biconical phase $XY$-$Z_{FIM}$ with ferrimagnetic $z$-order that does exist over a range of magnetic fields\,[7].

\section{Pseudospin description of the cuprate physics}
The pseudospin Hamiltonian, Eq.(\ref{H})  differs  from its  simplified version (\ref{Hs}) in several points.
This concerns the charge density constraint (\ref{Sz}), a significantly more complicated form of the "transversal"\, ($XY$) term, existence of the conventional spin s=1/2 for $M=0$ pseudospin states.   
The pseudospin parameters, in particular $\Delta$, $V_{ij}$, $h$ in the effective Hamiltonian (\ref{H}) can be closely linked to each other. 
At variance with typical spin systems the pseudospin system appears to be strongly anisotropic one with an enhanced role of frustrative effects of in-plane next-nearest neighbor couplings, inter-plane coupling, and different non-Heisenberg biquadratic interactions.
Despite the difference we can translate many results of the spin S\,=\,1 algebra to our pseudospin system. 
Turning to a classification of the possible homogeneous phases of the charge states of the model cuprates and its phase diagram we introduce $monovalent$  (Cu$^{1+}$, Cu$^{2+}$, Cu$^{3+}$), $bivalent$ MV-2 (Cu$^{1+,2+}$, Cu$^{2+,3+}$, Cu$^{1+,3+}$),  and $trivalent$ MV-3 (Cu$^{1+,2+,3+}$)  phases in accordance with character of the on-site superpositions (\ref{site}). 
Then, in accordance with the above nomenclature of spin phases  and  the charge triplet -- S\,=\,1 pseudospin correspondence we arrive at a  parent monovalent (Cu$^{2+}$) phase as an analogue of the QPM phase,  the $XY^{13}$,  $XY^{123}$, $XY$-$Z_{FM}^{13}$, $XY$-$Z_{FM}^{23}$, $XY$-$Z_{FM}^{12}$,  $XY$-$Z_{FM}^{123}$,  $XY$-$Z_{AFM}^{13}$,  $XY$-$Z_{AFM}^{123}$, $XY$-$Z_{FIM}^{13}$,  $XY$-$Z_{FIM}^{123}$,  $Z_{FM}^{1}$, and $Z_{FM}^{3}$  phases as mono-, bi-, and trivalent analogues of respective spin phases. All the metallic phases with $XY^{13}$ and  $XY^{123}$ components do admit in principle the  pseudospin nematic  order $\langle S_{\pm}^2\rangle\not=$\,0  related with  the high-T$_c$ superconductivity (HTSC). In all the trivalent phases the superconducting order competes with a spin ordering. Moreover, in $XY$-$Z_{FIM}^{123}$ phase we deal with a competition of superconducting, spin, and charge orders. 
It is worth noting that the $XY$-$Z$ nomenclature does strictly reflect an interplay of kinetic  ($XY$-terms) and potential ($Z$-term)  energies, or itineracy and localization. 

For the undoped model cuprate with $\sum_i\langle S_{iz}\rangle$\,=\,0 (half filling) given rather large positive $\Delta >\Delta_1$ we arrive at insulating monovalent {\it quantum paramagnetic} $M^0$ (Cu$^{2+}$)-phase, a typical one for Mott-Hubbard insulators.  
In parent cuprates, such as La$_2$CuO$_4$, the Cu$^{2+}$ ions
form an antiferromagnetically (AF) coupled square lattice of
s = 1/2 spins, which could possibly realize the resonant
valence bond (RVB) liquid of singlet spin pairs. In the RVB
state the large energy gain of the singlet pair state, resonating
between the many spatial pairing configurations, drives strong
quantum fluctuations נstrong enough to suppress long range
AF order.
However, by lowering the $\Delta$ below $\Delta_1$ the undoped cuprate can be  turned first into metallic and superconducting $XY^{123}$ phase, and given $\Delta <\Delta_2$ into a fully disproportionated MV-2 system of electron $M^-$ and hole $M^+$ centers ($M^{\pm}$-phase) with 
$\langle S_{iz}^2\rangle$\,=\,1, or {\it electron-hole Bose liquid} (EHBL)\,[1-4,8].
There is no single particle transport: $\langle S_{\pm}\rangle =\langle T_{\pm}\rangle $\,=\,0,  while the bosonic one may exist, and, in common, $\langle S_{\pm}^2\rangle \not=$\,0. 
Given the $\Delta \rightarrow -\infty$ condition the EHBL phase is equivalent to the lattice hard-core (hc) Bose system with an inter-site repulsion.
Indeed, one may  address the electron $M^-$ center  to be a system of a local composite  boson ($e^2$) localized on the hole  $M^+$ center: $M^- = M^+ + e^2$. 
 For such a system, the pseudo-spin Hamiltonian (\ref{H}) can be mapped
 onto the Hamiltonian of hc Bose gas on a lattice (see Refs.\,[9,10]  and references therein)
\begin{equation}
\smallskip H=\sum\limits_{i>j}t^b_{ij}(B_{i}^{\dagger}B_{j}+B_{j}^{\dagger}B_{i})
+\sum\limits_{i>j}V_{ij}N_{i}N_{j}-\mu \sum\limits_{i}N_{i},  \label{Bip}
\end{equation}
where $N_{i}=B_{i}^{\dagger}B_{i}$, ${\hat B}^{\dagger}({\hat B})$ (${\hat S}_{\pm}^2$) are
the Pauli creation (annihilation) operators which are Bose-like commuting for
different sites, but  $B_{i}^{2}=(B_{i}^{\dagger})^{2}=0$, $[B_{i},B_{i}^{\dagger}]=1-2N_i$. 
  The EHBL model exhibits many fascinating quantum phases and phase
transitions. Early investigations\,[9] point to the $T=0$ charge order (CO=$Z_{AFM}^{13}$), Bose
superfluid (BS=$XY$-$Z_{FM}^{13}$) and mixed (BS+CO=$XY$-$Z_{FIM}^{13}$) supersolid uniform phases with an Ising-type
melting transition (CO-NO=$Z_{AFM}^{13}$-$Z_{FM}^{13}$) and Kosterlitz-Thouless-type (BS-NO=$XY$-$Z_{FM}^{13}$-$Z_{FM}^{13}$) phase
transitions to a non-ordered normal fluid (NO=$Z_{FM}^{13}$) in 2D systems.
 At half-filling ($n_B=0.5, \Delta n=0$) given $t^b>V_{nn}$, $V_{nnn}$\,=\,0 the EHBL system obviously prefers a superconducting BS=$XY^{13}$ phase while at $t^b<V_{nn}$, $V_{nnn}$\,=\,0
it prefers an insulating checkerboard charge order CO=$Z_{AFM}^{13}$.
 It is worth noting that the QMC calculations\,[11] show that under doping away from half filling, the checkerboard solid undergoes phase separation: the superfluid (BS) and solid (CO) phases coexist but not as a single thermodynamic BS+CO phase. 

The EHBL model truly reproduces many important aspects of the cuprate physics\,[1], in particular, the pseudogap phenomenon as a result of the EH coupling. At the same time it cannot explain a number of well-known properties, in particular, manifestation of the Cu$^{2+}$ valence states in doped cuprates  over wide doping range\,[12] and suppression of the superconductivity for overdoped cuprates. Such a behaviour cannot be derived from the EHBL scenario and  points to realization of the more complicated "boson-fermion"\, dual XY-Z$_{FIM}^{123}$ phase with coexisting spin and pseudospin (charge) orders in a wide doping range from parent to  overdoped compounds including all  the  superconducting phase.
The suppression of the superconductivity for the hole overdoped cuprates can be explained as a transition from the trivalent superconducting (M$^{123}$) phase to a bivalent nonsuperconducting M$^{23}$ phase. Indeed, the M$^{-}$=Cu$^{1+}$ centers could be energetically gainless under hole doping particularly for overscreened EH coupling.
Some properties of nonsuperconducting phases M$^{23}$ and M$^{12}$, or $XY$-$Z_{FM}^{23}$ and $XY$-$Z_{FM}^{12}$,  can be understood if we address limiting insulating phases $M^{+}$ or $M^{-}$ ($Z_{FM}^1$ or $Z_{FM}^3$) with precisely $M^{+}$ or $M^{-}$-centers on each of the lattice sites. In frames of the pseudospin formalism these phases correspond to fully polarized ferromagnetic states with $S_z^{tot}$\,=\,$\pm$\,N, where N is the number of Cu sites. Interestingly, in frames of the pseudospin formalism  the "heavily overdoped" \,$XY$-$Z_{FM}^{23}$ and $XY$-$Z_{FM}^{12}$ phases with $x\approx$\,1 can be represented as ferromagnets where the charge constraint is realized through the occurrence of (1-x)N non-interacting pseudospin magnons ($\Delta S_z$\,=\,$\pm$\,1), that is Cu$^{2+}$ centers, obeying Fermi statistics due to s=1/2 conventional spin. These heavily overdoped cuprates could be addressed to be conventional Fermi liquids with a large Fermi surface.

\section{Conclusion}
The S=1 pseudospin formalism is shown to provide a conceptual framework for an in-depth understanding and a novel starting point for analytical and computational studies of high-T$_c$ superconductivity and other puzzles in cuprates.

\ack{The work was supported in part  by Ural Federal University in frames of the program 02.A03.21.0006 and  RFBR Grant No.  12-02-01039.}

\end{document}